\documentclass[prl,twocolumn,showpacs,preprintnumbers,amsmath,amssymb]{revtex4-1}
\usepackage{graphicx}
\usepackage{dcolumn}
\usepackage{bm}
\usepackage{color}

\begin{document}
\title{High-accuracy optical clock based on the octupole transition in ${}^{171}\text{Yb}^+$ }
\author{N. Huntemann}
\email{nils.huntemann@ptb.de}
\author{M. Okhapkin}
\author{B. Lipphardt}
\author{S. Weyers}
\author{Chr. Tamm}
\author{E. Peik}
\affiliation{Physikalisch-Technische Bundesanstalt, Bundesallee 100, 38116 Braunschweig, Germany}

\date{\today}

\begin{abstract}
We experimentally investigate an optical frequency standard based on the 467 nm (642 THz) electric-octupole reference transition $^2S_{1/2}(F=0)\rightarrow{}^2F_{7/2}(F=3)$ in a single trapped ${}^{171}\text{Yb}^+$ ion. The extraordinary features of this transition result from the long natural lifetime and from the $4f^{13}6s^2$ configuration of the upper state. The electric-quadrupole moment of the $^2F_{7/2}$ state is measured as $-0.041(5)~ea^2_0$, where $e$ is the elementary charge and $a_0$ the Bohr radius. We also obtain information on the differential scalar and tensorial components of the static polarizability and of the probe-light-induced ac Stark shift of the octupole transition. With a real-time extrapolation scheme that eliminates this shift, the unperturbed transition frequency is realized with a fractional uncertainty of $7.1\times10^{-17}$. The frequency is measured as $642\ 121\ 496\ 772\ 645.15(52)~\text{Hz}$.
\end{abstract}


\maketitle

The basis of all precise atomic clocks is a transition frequency that represents an unperturbed quantum property of the chosen atomic system. The most impressive progress in clocks of high accuracy has recently been made with optical transitions between states with vanishing electronic angular momentum ($J=0$) in Al$^+$ and Sr \cite{Chou2010,Katori2011}. The frequency of this type of transition is in general only weakly affected by external electric and magnetic fields. Here we present a precision study of a reference transition of a very different type, an electric-octupole transition ($\Delta J=3$)  connecting the $^2S_{1/2}$ ground state with the $^2F_{7/2}$ first excited state in $^{171}\text{Yb}^+$, and show that it has a very low sensitivity to field-induced frequency shifts, making it a promising basis for an optical clock of the highest accuracy. 

At variance with other ion frequency standards, $^{171}\text{Yb}^+$ offers the advantage of two optical reference transitions with high quality factor which have rather different physical characteristics. A frequency standard based on the electric-quadrupole $^2S_{1/2}\rightarrow{}	^2D_{3/2}$ transition \cite{Schneider2005,Tamm2009} is established as one of the secondary representations of the SI second. The electric-octupole $^2S_{1/2} \rightarrow{}^2F_{7/2}$ transition investigated in this Letter was first studied at the National Physical Laboratory (UK) \cite{Roberts1997}. The extraordinary features of this transition result from the long natural lifetime of the $^2F_{7/2}$ state in the range of several years \cite{Roberts1997,Biemont1998} and from its electronic configuration $(4f^{13}6s^2)$ consisting of a hole in the $4f$ shell surrounded by a spherically symmetric $6s$ shell. Since the octupole transition can be resolved with a linewidth that is virtually unaffected by spontaneous decay and determined only by the available laser stability, a quantum projection noise limited single-ion frequency standard with very low  instability can be realized. The electric-quadrupole moment of the $^2F_{7/2}$ state is predicted to be much smaller than that of the $^2D_{3/2}$ state \cite{Blythe2003} so that the transition frequency is only weakly affected by the quadrupole shift from electric field gradients. Furthermore, there are no strong dipole transitions from the $^2F_{7/2}$ state with excitation energies below 3~eV. Therefore, one expects its quadratic Stark shift induced by static or infrared electric fields to be small. The important relativistic contributions to the $^2F_{7/2}$ state energy lead to a particularly strong sensitivity of the transition frequency to variations of the fine structure constant $\alpha$ \cite{Dzuba2009}. Measurements of the frequency ratio of the quadrupole and the octupole transitions in one trapped $^{171}\text{Yb}^+$ ion would be a convenient way to test the constancy of $\alpha$.

Because of the extremely small oscillator strength of the octupole transition, its excitation requires particularly high spectral power density. The required intensity leads to nonresonant couplings to higher lying levels and thereby introduces a significant light shift of the transition frequency. This strong light shift and the difficulty to efficiently excite the octupole transition have so far impeded the realization of an optical frequency standard and detailed investigations of systematic shifts, which are both presented in this Letter.
\begin{figure}
\includegraphics[width=.8\columnwidth]{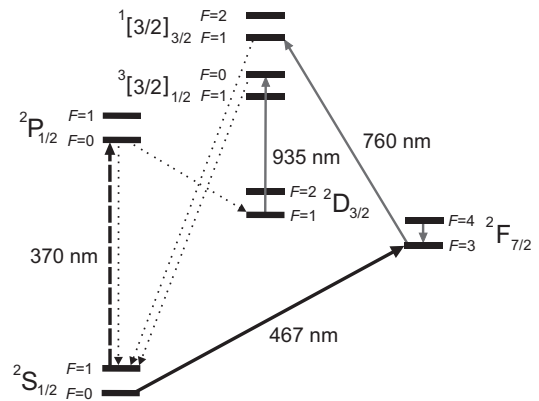}
\caption{Partial level scheme of $^{171}\text{Yb}^+$. The solid black arrow indicates the octupole transition under investigation. The cooling transition is shown by the dashed arrow, and solid gray arrows represent repumping transitions. Relevant spontaneous decay paths are indicated by dotted arrows. 
\label{LevelScheme}}
\end{figure}
In our experiment, a single ${}^{171}\text{Yb}^+$ ion is confined in a cylindrically symmetric radio frequency Paul trap \cite{Paul1990} and is laser-cooled by excitation of the ${}^2S_{1/2}\to{}^2P_{1/2}$ transition (see Fig.~\ref{LevelScheme}) at 370~nm \cite{Tamm2009}. Each measurement cycle starts with a cooling period ($10$~ms) followed by hyperfine pumping to the $^2S_{1/2} (F=0)$ ground state ($25$~ms). Subsequently, the cooling and repumping lasers are blocked and a rectangular 467~nm probe pulse is applied. Absence of the fluorescence at the beginning of the following cooling period indicates excitation of the $^2F_{7/2}$ state. After the detection period ($3$~ms), a 760~nm repumping excitation returns the ion to the ground state for laser cooling. This new repumping scheme is more efficient than the previously used scheme \cite{Gill1995} and makes use of the electric-quadrupole transition to the $^1[3/2]_{3/2} (F=1)$ state at $34575~\text{cm}^{-1}$, which has a natural lifetime of 28.6~ns \cite{Berends1993} and predominantly decays to the ground state. With a laser power in the milliwatt range, the $^2F_{7/2}$ state is depleted in a few milliseconds. In the infrequent case that the $^2F_{7/2} (F=4)$ state is populated, the hyperfine transition to $F=3$ is driven by microwave radiation at 3.6~GHz (see Fig.~\ref{LevelScheme}).

The octupole transition is driven by a probe laser system with a relative frequency stability of better than $2\times10^{-15}$ at 1~s of averaging time \cite{Sherstov2010}. The frequency-doubled laser output at 467~nm is coupled into a blue-emitting laser diode for injection locking. This scheme minimizes intensity fluctuations that can lead to significant line broadening caused by a fluctuating light shift. A power of up to 12~mW is focused to a beam waist diameter of $\approx$40~$\mu$m at trap center. During the probe pulses, a magnetic field of 3.60(2)~$\mu$T is applied in order to separate the Zeeman components. Its strength and orientation is inferred from spectroscopy of the $^2S_{1/2}\rightarrow{} ^2D_{3/2}$ transition \cite{Tamm2009}. The laser intensity incident  on the ion is monitored with a photodetector mounted behind the trap. A pinhole in front of the photodetector selects the center of the magnified image of the laser beam waist at the position of the ion so that intensity fluctuations due to both total power and pointing instabilities are detected. The photodetector uses a Si \textit{p-n} photodiode with a very low photocurrent nonlinearity in the $10^{-5}$ range \cite{LeiFischer1993}. The high linearity enables the precise measurement of intensity ratios that is required for eliminating the light shift (see below).   

Figure~\ref{Spectrum} shows the excitation spectrum of the $(F=0)\rightarrow (F=3,m_F=0)$ component of the octupole transition. With a probe pulse duration of 120~ms and a focused laser power of about 0.5 mW, we obtain a Fourier-limited linewidth of 6.6~Hz and a resonant excitation probability of more than 90\%. For longer pulse durations, the line shape of the resonance signal is degraded through the frequency instability of the probe laser that is determined by the thermomechanical noise of the reference cavity \cite{Sherstov2010}. The light shift present in the spectrum shown in Fig.~\ref{Spectrum} is significantly larger than the linewidth and amounts to $28$~Hz. Generally, one expects that the light shift contains both scalar and tensorial contributions and that the shift $\Delta\nu$ caused by a $\pi$-pulse with Fourier-limited spectral width $\Delta f$ is proportional to $(\Delta f)^2$. By recording excitation spectra for various probe pulse areas and durations, we find $\Delta\nu = 0.65(3)$~Hz$^{-1}(\Delta f)^2$ if the polarization and magnetic field orientation are chosen to maximize the excitation probability. The measurement uncertainty here arises mainly from changes in the probe laser frequency drift rate.

\begin{figure}
\includegraphics[width=\columnwidth]{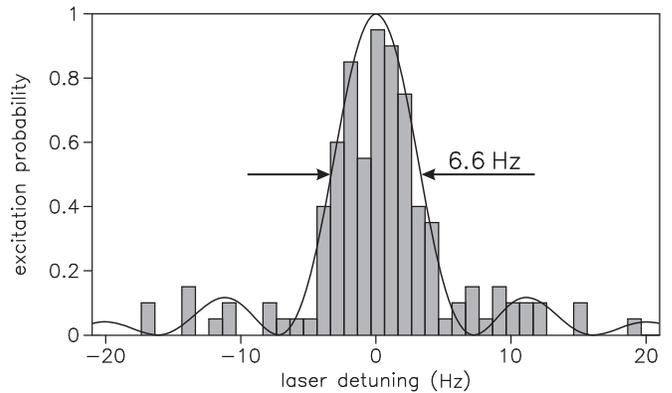}
\caption{Excitation spectrum of the $^2S_{1/2} (F=0)\to{}^2F_{7/2}(F=3,m_F=0)$ transition obtained with a probe pulse duration of 120~ms. For each laser detuning step, 20 measurement cycles were performed. The solid line shows the theoretical line shape for $\pi$-pulse excitation.\label{Spectrum}}
\end{figure}

Making use of the high spectral resolution and excitation efficiency, we investigate the frequency shifts due to external electric fields. For this, we use an interleaved servo technique: the parameter causing the shift is switched between two settings every fourth measurement cycle and the probe light frequency is stabilized to the respective line centers \cite{Tamm2009}. The method allows us to measure systematic frequency shifts with uncertainties in the millihertz range. 

\begin{figure}
\includegraphics[width=.9\columnwidth]{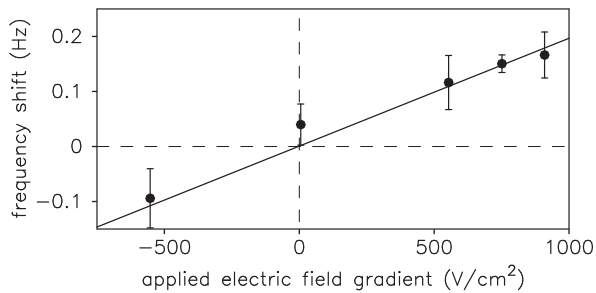}
\caption{Quadrupole shift of the octupole transition frequency resulting from an applied electric field gradient, measured using an interleaved servo technique (see text). The solid line is a linear least-square fit through the origin. The error bars denote statistical uncertainty. \label{QuadShift}}
\end{figure}

The interaction of the electric-quadrupole moment $\Theta(F,7/2)$ of the $^2F_{7/2}$ state with a static electric field gradient leads to a frequency shift of the octupole transition. Significant shifts can be caused, in particular, by the uncompensated gradient of stray fields in the trap. In our case, the quadrupole shift produced by a rotationally symmetric electric field gradient is given by $\Delta\nu_Q=\frac{5}{7h}A\,\Theta(F,7/2)\,(3\cos^2\beta-1)$ \citep{Itano2000}, where $h$ is the Planck constant and $\beta$ is the angle between the symmetry axis of the gradient and the applied magnetic field. To determine $\Theta(F,7/2)$, a known electric field gradient of strength $A=U_\text{dc}/d^2_0$ is produced by applying a static voltage $U_\text{dc}$ to the trap ring electrode.  The geometrical factor $d_0=1.1181(4)~\text{mm}$ was determined by fitting measurements of the secular frequencies for different dc voltages to a theoretical model, as in Ref.~\cite{Barwood2004}. Using the interleaved servo technique with $U_\text{dc}$ alternating between zero and a preset value, the measurement shown in Fig.~\ref{QuadShift} was performed with $\beta=90(1)^\circ$. The slope of the fit yields $\Theta(F,7/2)=-0.041(5)$~$ea^2_0$, where $e$ is the elementary charge and $a_0$ the Bohr radius. The uncertainty is dominated by the statistical uncertainty of the measured frequency differences. This experimental value is significantly lower than a previous theoretical estimate, $\Theta(F,7/2)=-0.22$~$ea^2_0$, obtained by a single-configuration Hartree-Fock calculation \cite{Blythe2003}. The discrepancy is not surprising in view of the complex electronic configuration of the $^2F_{7/2}$ state. 

It appears that so far no experimental data on the static differential polarizability of the octupole transition are available. The frequency shift $\Delta\nu_\text{QS}$ of the investigated transition due to the static quadratic Stark (QS) effect can be expressed as $\Delta\nu_\text{QS}=\left[2\Delta\alpha^\text{dc}_s+\frac{4}{5}\alpha^\text{dc}_t(3\cos^2\theta-1)\right]E^2/(4h)$, where $\Delta\alpha^\text{dc}_s=\alpha^\text{dc}(^2S_{1/2})-\alpha^\text{dc}(^2F_{7/2})$ is the difference in the static polarizabilities of the ground state and the excited state, $\alpha^\text{dc}_t$ is the static tensor polarizability of the excited state, and $\theta$ is the angle between the electric field and the magnetic field \cite{Itano2000}. A previously applied technique to determine the polarizability of a trapped ion uses a displacement from trap center to expose the ion to the rf trap field \cite{Yu1994,Schneider2005}. In our case, the displacement within the intensity profile of the probe laser would cause an unacceptably large uncertainty due to the change in light shift. To avoid this problem, we apply a fraction of the ac trap drive voltage with a phase shift of $\pi/2$ to one of the end caps of the trap. The additional field induces micromotion of the ion but does not lead to displacements that affect the light shift. From the measured frequency shift, we find $\Delta\alpha^\text{dc}_s=-1.3(6)\times10^{-40}$~J V$^{-2}$m$^2$. By changing the orientation of the applied magnetic field, we find that the tensor polarizability $\alpha^\text{dc}_t$ is approximately 1 order of magnitude smaller than $\Delta\alpha^\text{dc}_s$. The uncertainty in $\Delta\alpha^\text{dc}_s$ results from the uncertainty in the actual electric field strength acting on the ion and from the correction for the relativistic Doppler effect, which contributes significantly to the measured shift. Our result is in good agreement with calculations \cite{Lea2006}. In another measurement, we determined the ratio between the tensorial and scalar dynamic polarizabilities at the probe light frequency as $\alpha^{ac}_t(467~\text{nm})/\Delta\alpha^{ac}_s(467~\text{nm})= -0.137(5)$.

\begin{figure}
\includegraphics[width=\columnwidth]{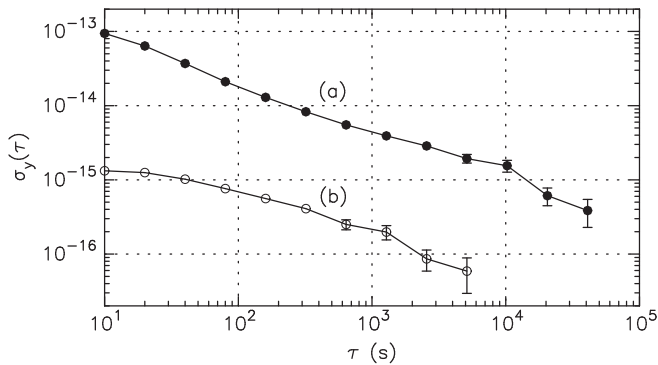}
\caption{Fractional Allan deviation $\sigma_y(\tau)$ of (a) the absolute frequency measurement versus the fountain clock CSF1 and (b) the light shift induced offset frequency $\nu_\text{offset}$ (see text) in measurement 5 in Table~\ref{AbsValues}. \label{AllanDev}}
\end{figure}

In previous investigations, the absolute frequency of the octupole transition was measured for different probe light powers and then linearly extrapolated to zero power to correct for the light shift \cite{Hosaka2009}. We use the interleaved servo technique to obtain the unperturbed frequency by real-time extrapolation. By changing the probe light power, the intensity is alternated between a value $I_H$ and a lower value  $I_L$ and the laser frequency is stabilized to the corresponding light-shifted line center frequencies $\nu_H$ and $\nu_L= \nu_H - \nu_\text{offset}$ \cite{Peik2006}. The unperturbed transition frequency is calculated as $\nu_0 = \nu_H - \nu_\text{offset}\left(1-I_L/I_H\right)^{-1}$ so that it depends not on the absolute values of the probe pulse intensities but on their ratio, which is registered together with $\nu_H$ and $\nu_\text{offset}$. In this way the frequency of the octupole transition was measured with a fiber-laser-based frequency comb generator \cite{Lipphardt2009} using the caesium fountain clock CSF1 \cite{Weyers2001} in our laboratory as the reference.  Table~\ref{AbsValues} shows the results of eight measurements with different settings of $I_H$ and $I_L$. All results agree within the statistical uncertainty which is determined by the frequency instability of the caesium fountain clock \cite{Tamm2009} and yield no evidence for systematic shifts related to the choice of $I_H$ and $I_L$. Figure~\ref{AllanDev}(a) shows the combined Allan deviation of all measurements listed in Table~\ref{AbsValues} covering a total measurement time of 87~h. In order to illustrate the stability of the light shift, Fig.~\ref{AllanDev} also shows the Allan deviation of $\nu_\text{offset}$ for measurement 5 in Table~\ref{AbsValues} which was conducted with a particularly high shift. The instability of $\nu_\text{offset}$ is consistent with the expected level of quantum projection noise reaching $\sigma_y(5000~s)=6\times10^{-17}$. 

\begin{table}
\caption{Deviations $\Delta\nu=\nu-\nu_0$ of frequency measurements of the octupole transition from the mean value $\nu_0=642121496772645.15 (52)~\text{Hz}$. The measurements were performed with various combinations of probe pulse intensities and resulting time-averaged light shift magnitudes $\Delta\nu_H$ and $\Delta\nu_L$. The intensity ratios recorded during the measurements were used in an interleaved servo extrapolation scheme (see text). $t$: measurement time, $\sigma^{\text{m}}_\nu$: statistical uncertainty. \label{AbsValues}}

\begin{ruledtabular}
\begin{tabular} { r r r r r r }
 No.  &   $\Delta\nu_H$~(Hz) & $\Delta\nu_L$~(Hz) &  $t$~(h)   &   $\Delta\nu$~(Hz) & $\sigma^{\text{m}}_\nu$~(Hz) \\ \hline 
 1    &   $194$        & $36$         &  $5.2$     &   $0.29$             &    $0.63$         \\ 
 2    &   $196$        & $32$         &  $11.0$    &   $-0.22$            &    $0.46$         \\    
 3    &   $199$        & $33$         &  $2.9$     &   $0.53$             &    $0.96$         \\    
 4    &   $202$        & $34$         &  $10.0$    &   $0.06$             &    $0.52$         \\  
 5    &   $412$        & $59$         &  $23.7$    &   $0.34$             &    $0.26$         \\ 
 6    &   $401$        & $62$         &  $19.2$    &   $-0.62$            &    $0.31$         \\    
 7    &    $71$        & $42$         &  $7.8$     &   $0.38$             &    $0.49$         \\ 
 8    &   $160$        & $118$        &  $7.1$     &   $-0.27$            &    $0.44$             
\end{tabular}
\end{ruledtabular}
\end{table}

The mean of the measurements summarized in Table~\ref{AbsValues} determines the  frequency of the unperturbed $^2S_{1/2}(F=0)-{}^2F_{7/2}(F=3,m_F=0)$ transition to be $642\ 121\ 496\ 772\ 645.15 (52)~\text{Hz}$. Here, the total fractional uncertainty of $8.0\times10^{-16}$ has a statistical contribution of $2.5\times10^{-16}$, a systematic contribution due to the fountain clock of $7.6\times10^{-16}$, and a systematic uncertainty contribution of $7.1\times10^{-17}$  due to perturbations of the octupole transition frequency (see below). The achieved total uncertainty represents an improvement by more than a factor of 20 in comparison with the previously published result \cite{Hosaka2009}.  
 
Since the dominant shift of the octupole transition frequency is the light shift, it is essential to quantify the uncertainty of the employed extrapolation scheme. We expect that the nonlinearity of the intensity measurement and higher-order terms in the light shift do not contribute significantly to this uncertainty. We also expect that the registered average of $I_L/I_H$ reliably represents the intensity ratio at the position of the ion and that probe pulse intensity fluctuations are not correlated with the execution of the measurement cycles. The observed single-pulse intensities $I_L$ and $I_H$ show relative fluctuations of $\approx$1\% with a slightly asymmetric distribution presumably resulting from pointing instability. Under the assumptions made here, these fluctuations will not lead to an error in the extrapolated frequency and would result in a symmetric distribution of $I_L/I_H$. The observed distributions, however, show a small asymmetry which we use as a measure for the extrapolation uncertainty. We quantify the asymmetry as the difference between the mean and the most probable value of $I_L/I_H$. The corresponding deviations in the extrapolated frequencies for the measurements listed in Table~\ref{AbsValues} are in the range -27 \dots +19~mHz and we use the maximum absolute deviation as an estimate for the extrapolation uncertainty. 

Table~\ref{UncertCon} lists the leading systematic frequency shifts and uncertainty contributions present during the measurement of the octupole transition frequency. The frequency shift due to blackbody radiation can be calculated from the measured static polarizability $\Delta\alpha^\text{dc}_s$. The static description is a good approximation here because there are no strong infrared dipole transitions originating from the $^2S_{1/2}$ and the $^2F_{7/2}$ state. We obtain a shift of $-0.067(32)$~Hz at $T=300$~K. 

The uncertainty contribution from the quadrupole shift results from the measured very small quadrupole moment of the $^2F_{7/2}$ state and the maximum stray field gradient previously observed in the same trap \citep{Tamm2009}. A stray field compensation as described in Ref. \cite{Schneider2005} leads to the listed uncertainty contribution for the second-order Doppler shift and determines, together with the measured static differential polarizability, the uncertainty contribution due to the quadratic dc Stark shift. The low nonlinear frequency drift of the probe laser \cite{Sherstov2010} and the use of a second-order integrating servo scheme for the stabilization to the atomic resonance \cite{Peik2006} leads to the estimated uncertainty due to servo error. The small sensitivity to the quadratic Zeeman effect \cite{Hosaka2005} results in a negligible uncertainty contribution.

\begin{table}
\caption{Leading fractional shifts $\delta\nu/\nu_0$ of the octupole transition frequency $\nu_0$ and uncertainty  contributions $u/\nu_0$.  \label{UncertCon}}
\begin{ruledtabular}
\begin{tabular} { l c c c}
 Effect                           & $\delta\nu/\nu_0~(10^{-18})$                & $u/\nu_0~(10^{-18})$  \\ \hline 
 Blackbody radiation shift        & $-105$                                      & $50$    \\
 Light shift extrapolation        & $0$                                         & $42$    \\ 
 Quadrupole shift                 & $0$                                         & $22$    \\ 
 Second-order Doppler shift       & $0$                                         & $16$    \\ 
 Quadratic dc Stark shift         & $0$                                         & $4$     \\   
 Servo error                      & $0$                                         & $3$     \\
 Second-order Zeeman shift        & $-36$                                       & $1$     \\ \hline
 Total			                      &	$-141$	                                    & $71$  				
\end{tabular}
\end{ruledtabular}
\end{table}

We have demonstrated that the $^2S_{1/2}\to{}^2F_{7/2}$ octupole transition in ${}^{171}\text{Yb}^+$ can be used to realize an optical clock with a systematic uncertainty of $7.1\times10^{-17}$. It appears straightforward to reduce the uncertainty further with the use of established techniques that minimize shifts caused by the electric stray field \cite{Itano2000,Berkeland1998}. The presently dominant systematic uncertainty contribution due to the blackbody radiation shift can be reduced by a more precise experimental determination of the static scalar differential electric polarizability or by improved atomic structure calculations. The latter can take advantage of the measurements of atomic parameters presented here. Moreover it has recently been shown that a clock based on a linear combination of the quadrupole and the octupole transition frequencies of Yb$^+$ can have a significantly reduced blackbody shift \cite{Yudin2011}. Finally, a novel multipulse Ramsey excitation scheme promises to suppress the light shift by the probe radiation by at least 2 orders of magnitude \cite{Yudin2010}. 

We thank I. Sherstov for his contribution to the construction of the probe laser system. This work was supported by the DFG through QUEST.

\textit{Note added in proof.}---Since the submission of this paper, the result of an independent measurement of the absolute frequency of the octupole transition has been published \cite{King2012}.  It is in good agreement with the value presented here.

\end{document}